# Electron diffraction covering a wide angular range from Bragg diffraction to small-angle diffraction

Hiroshi Nakajima[1], Atsuhiro Kotani[1], Ken Harada[1, 2], and Shigeo Mori[1, *]

[1]*Department of Materials Science, Osaka Prefecture University, Sakai, Osaka 599-8531, Japan*

[2]*Center for Emergent Matter Science, the Institute of Physical and Chemical Research (RIKEN), Hatoyama, Saitama 350-0395, Japan*

*To whom correspondence should be addressed. E-mail: mori@mtr.osakafu-u.ac.jp

We construct an electron optical system to investigate Bragg diffraction (the crystal lattice plane, $10^{-2}$–$10^{-3}$ rad) with the objective lens turned off by adjusting the current in the intermediate lenses. A crossover was located on the selected-area aperture plane. Thus, the dark-field imaging can be performed by using a selected-area aperture to select Bragg diffraction spots. The camera length can be controlled in the range of 0.8 to 4 m without exciting the objective lens. Furthermore, we can observe the magnetic-field dependence of electron diffraction using the objective lens under weak excitation conditions. The diffraction mode for Bragg diffraction can be easily switched to a small-angle electron diffraction mode having a camera length of more than 100 m. We propose this experimental method to acquire electron diffraction patterns that depict an extensive angular range from $10^{-2}$ to $10^{-7}$ rad. This method is applied to analyze the magnetic microstructures in three distinct magnetic materials, i.e., a uniaxial magnetic structure of $BaFe_{10.35}Sc_{1.6}Mg_{0.05}O_{19}$, a martensite of a Ni–Mn–Ga alloy, and a helical magnetic structure of $Ba_{0.5}Sr_{1.5}Zn_2Fe_{12}O_{22}$.

**Introduction**

Electron diffraction is one of the most effective methods to examine various physical phenomena and phase transitions in functional materials. Various methods are utilized to perform electron diffraction in transmission electron microscopy. A diffraction pattern from a micrometer- or a nanometer-scaled area can be obtained using selected-area electron diffraction (SAED). A real space image, such as a bright or a dark-field image, is obtained to depict the morphology of domain structures in the same area as that of the SAED pattern [1, 2]. Furthermore, nano-beam electron diffraction (NBED) is applied to analyze the local structures using a coherent electron beam having a diameter that is smaller than 1 nm [3]. Additionally, convergent-beam electron diffraction (CBED) is used to measure the strain and to determine a space group unambiguously owing to the dynamic diffraction effect in a nanometer-scaled area [4]. Diffraction patterns obtained using NBED and CBED provide information about the local crystallographic structures.

Small-angle electron diffraction (SmAED) that uses a long camera length having a length in the range of several tens or hundreds of meters is another useful technique in electron diffraction [5, 6]. In ferromagnetic materials, the angles of magnetic deflection and that of diffraction from long wavelength periodic domains are observed to be smaller than those of the crystallographic Bragg diffraction by two or three orders of magnitude. In typical ferromagnets, the Bragg diffraction angles from the crystal lattice planes are $10^{-2}$–$10^{-3}$ rad when the accelerating voltage of the electrons is 200 kV. Simultaneously, the magnetic deflection angles are observed to be $10^{-4}$–$10^{-6}$ rad [7, 8]. Therefore, SmAED is required to disclose the magnetic domain structures as well as to measure the magnetization magnitude. More recently, SmAED has been utilized to analyze the magnetic domain structures and the magnetic and electric fields of various magnetic materials and devices [9–14]. In our previous studies [15, 16], we constructed an electron optical system having a long camera length under controlled magnetic fields for both Foucault method and SmAED. This optical system further makes it possible to use a camera length of up to approximately 1300 m.

In addition to the SmAED patterns, the observations of Bragg diffraction are important for some magnetic materials, such as transition metal oxides and martensites, because their crystallographic domains are related with their magnetic domains. In such materials, the dark-field imaging method is effective to study the crystallographic domains. Additionally, because the magnetization directions depend on the crystallographic axes, tilting a specimen based on the diffraction patterns is required for observing the magnetic domain. The magnetic field that is created by the objective lens induces magnetostriction and alters the magnetic domain structures. Therefore, turning off the objective lens is necessary to perform these observations. In the SmAED optical system [15, 16], however, Bragg diffraction spots are blocked by the fixed apertures in the imaging system because Bragg diffraction angles are larger than those of the magnetic deflection, which makes it impossible to obtain Bragg diffraction patterns. Furthermore, the commercially available optical systems that are preinstalled by the manufacturers cannot provide diffraction patterns and dark-field images while using a conventional transmission electron microscope (TEM) when the objective lens is turned off.

In this study, we constructed an optical system for observing both the Bragg diffraction and magnetic deflection in the same area. We used three intermediate lenses to form a compound lens in order to reduce the camera length to be as short as 1 m. The illumination system in this optical system was identical to that in SmAED. Thus, we can easily switch the electron diffraction optical system to form an SmAED optical system by altering the current in the intermediate lenses. We applied this optical system to observe both the crystallographic and magnetic domains as well as the Bragg diffraction and magnetic deflection patterns in uniaxial $BaFe_{10.35}Sc_{1.6}Mg_{0.05}O_{19}$ (BFSMO) [17, 18] and a ferromagnetic Ni–Mn–Ga martensite (NMG) [19]. We additionally observed a magnetic-field induced phase transition (an alteration in the magnetic modulation vectors of the helical structures) in the Y-type hexaferrite $Ba_{0.5}Sr_{1.5}Zn_2Fe_{12}O_{22}$ (BSZFO) [20].

**Methods**

An electron optical system for observing real-space images and Bragg diffraction was constructed using a conventional 200 kV TEM (JEM-2010, JEOL Ltd.). This electron microscope comprised an objective mini-lens, three intermediate lenses (first, second, and third intermediate lenses), and one projection lens in the imaging system. Thin specimens were prepared by Ar-ion milling for BFSMO and BSZFO and by jet electropolishing for NMG. The specimens of NMG and BSZFO were cooled using a liquid nitrogen-cooling holder.

**Results and discussion**

**Optical system**

Figure 1 depicts an electron optical system that was constructed for observing Bragg diffraction. This optical system is similar to the SmAED optical system in two manners [15, 16]. The condenser lens and objective mini-lens were adjusted to construct the crossover on the selected-area (SA) aperture plane. The objective lens was turned off for observing the magnetic materials and for applying controlled external magnetic fields. However, in the SmAED optical system, the first intermediate lens was adjusted to focus the real image (or crossover). The second and third intermediate lenses were used to modify the camera length. Conversely, in the optical system for Bragg diffraction, the current of the first intermediate lens was fixed to be lower than that in the SmAED mode to prevent Bragg diffraction spots from being blocked by the fixed apertures, which were located both below and above the projection lens. The second and third intermediate lenses were adjusted to focus the specimen image (or diffraction pattern) and to alter the camera length. Thus, the diffraction [Fig. 1(a)] and real-image [Fig. 1(b)] modes in the constructed optical system can be switched only by altering the currents in the second and third lenses. By selecting a Bragg diffraction spot using the SA aperture, the dark-field imaging can be performed without the presence of external magnetic fields of the objective lens. Furthermore, the electron diffraction mode can be switched to the SmAED mode by modifying the current in the intermediate lenses. Additionally, the Foucault method can also be performed by altering the current in the intermediate lenses after selecting the magnetically deflected spots. In the electron diffraction (SmAED) and dark-field modes, the illumination system was identical to that in the usual observation modes. However, the condenser lens was strongly excited to produce the small crossover spot in the SmAED mode.

We measured the camera lengths of the constructed optical system. Figure 2 depicts the current dependence of the camera length. The camera length in this optical system ($I_1 = 2.3$ A) was shorter as compared with that in the SmAED mode ($I_1 = 5.0$ A) and ranged from 4.5 to 0.8 m, which made it possible to observe Bragg diffraction using crystalline materials. The current in the second intermediate lens ($I_2$) was reduced to fix the crossover on the SA-aperture plane if the current in the third intermediate lens ($I_3$) was increased. In the SmAED mode, we controlled the focus using the first intermediate lens. Thus, the value of $I_3$ was not limited to alter the camera length when $I_2$ was fixed to 6 A. The dependence of $I_3$ was linear with respect to $I_2$. As indicated in Fig. 2, this method covered an extensive range of camera lengths (reciprocal space) that is sufficient to examine the crystallographic and magnetic domains.

**Magnetic domain observations using the constructed optical system**

We applied this optical system to observe the domains and phase transitions in the magnetic materials. Figures 3(a) and 3(b) show the dark-field image and diffraction pattern of BFSMO. Figure 3(a) was captured at defocus to image the magnetic domain walls. Figure 3(b) demonstrates that Bragg diffraction can be obtained without exciting the objective lens. This diffraction pattern depicted a six-fold symmetry, which resulted from the hexagonal crystal structure of BFSMO. The dark-field image depicted broad bright and dark contrast that originated from the thickness change or bend of the specimen and the striped contrast of the magnetic domain walls. The striped domain walls were caused by the Bloch walls that separated the magnetic domains [18]. The existence of Bloch walls can be confirmed by the streaks in the SmAED pattern [see the inset of Fig. 3(b)]. The spots in this SmAED pattern (blue arrows) indicated an angle $\theta$ of approximately 6.7 μrad, which corresponded to the periodicity $d = \lambda/\theta \sim 373$ nm ($\lambda = 2.51$ pm is the wavelength of the electron). This result agrees with twice the period of the striped domains (~ 180 nm) that was obtained from the real image [see the inset of Fig. 3(a)]. We consider that the specimen was slightly tilted and that the perpendicular magnetizations contained in-plane components, which caused diffraction spots that depict double periodicity as confirmed using another ferromagnet [21]. We note that the magnetic deflection was superimposed on Bragg diffraction spots as well as the direct beam spot 000 [22]. Information about the magnetic domain structure can be obtained from a dark-field image using the Bragg diffraction spot. Thus, if magnetic domains are correlated with the specimen thickness, the relation can be observed in a dark-field image at defocus. The width of the striped magnetic domains that were depicted in Fig. 3(a) did not change at the observed area. Conversely, the thickness fringes originating from the thickness change were extensively distributed. Therefore, the magnetic domains were observed to be populated without being disturbed by the thickness change as shown in Fig. 3.

Further, we observed NMG to demonstrate that this optical system was useful to reveal the relation between crystallographic and magnetic domains. The Fresnel image in Fig. 4(a) depicts the bright and dark lines originating from the magnetic domain walls and the discontinuities of bend contour indicating the crystallographic domains. These domain characters were clearly visualized in the dark-field [Fig. 4(b)] and Foucault [Fig. 4(c)] images. In the Bragg diffraction pattern that is depicted in the inset of Fig. 4(b), one fundamental spot was divided into two peaks. By selecting the peak that was indicated using the arrowhead, crystallographic twin domains were illustrated to be domains with bright and dark contrast in Fig. 4(b). These twin domains are called variants in martensite alloys and these variants are important for understanding magnetostriction [23, 24]. In the magnetic deflection pattern that is presented in the inset of Fig. 4(c), the direct beam 000 is split into four spots, which indicates that the magnetization had four directions. In this experiment, the camera length was altered from Bragg diffraction to the magnetic deflection pattern by varying the current in the intermediate lenses, as explained previously. By choosing the two spots (blue and red arrowheads) using an SA aperture, the magnetic domains containing the magnetization of the two directions (blue and red arrows) were visualized to depict bright contrast. Comparing Figs. 4(b) and 4(c), 180° magnetic domain walls were located in the same variants, and 90° domain walls were formed between different variants.

These results demonstrate that crystallographic and magnetic domains can be observed in the same area using the current optical system. Although the magnetic domains (Fresnel and Foucault images) and variants (dark-field images) in martensitic alloys are observed using TEMs [25–27], the observations of magnetic and crystallographic domains in the same area are rare, which may be caused due to the difficulty in observation while using preinstalled optical systems, as explained in the introduction section. We further note that variants are reported to be controlled by magnetic fields, which are the origins of the large magnetoelastic effect [28]. Thus, observing the dynamic behaviors of variants by external magnetic fields becomes possible using this optical system.

We estimated the magnetic flux density induced by the magnetization using this SmAED pattern. The half angle of the split spots was $\beta = \sim 3.5 \times 10^{-5}$ rad. Based on the equation $\beta = e\lambda Bt/h$ [5], this value corresponded to the magnetic flux density $B = \sim 0.72$ T when the thickness was assumed to be $t = 80$ nm. Here, $e$ is the electric charge, and $h$ is the Planck constant. This magnetic flux density was comparable to the values that were determined using the magnetization measurements in the Ni–Mn–Ga alloys [24, 25]. We note that the midpoint between the red and yellow spots did not coincide with that between the blue and green spots in the SmAED pattern. This may be caused due to the astigmatism of the lens or the spin configuration, i.e., the red and yellow magnetizations were canted and their angles were deviated from 180°, which was inferred using the curved domain wall between the (red and yellow) magnetic domains.

Finally, we observed BSZFO to demonstrate that magnetic-field induced phase transitions can be observed through diffraction using this optical system. Figure 5(a) depicts the diffraction pattern in BSZFO in the absence of an external magnetic field. This pattern depicted weak superlattice spots from a magnetic structure in addition to the Bragg diffraction peaks from a crystal structure [see red arrowheads in Fig. 5(b)]. When we applied the magnetic fields, a phase transition occurred, which was confirmed by the alterations in the superlattice spots in Figs 5(c) and 5(d). The magnetic field was applied parallel to the optical axis using the objective lens. It is considered that this phase transition originates due to the magnetic structures varying from the proper helical to transverse conical structures [Fig. 5(e)], which also plays an important role in the magnetoelectric effect [29]. We observed the diffraction pattern at other magnetic fields, and the magnetic-field dependence of the superlattice spots is depicted in Fig. 5(f). The superlattice spot ($\delta$) increased from 0.23 with the increasing magnetic field, and a $\delta$ of ~ 0.5 was also observed when $\delta$ was 0.25. This magnetic-field dependence is similar to that observed in a previous study [30], which further shows the effectiveness of the constructed optical system. As demonstrated using the three magnetic materials, this electron optical system is useful to study the relations between crystallographic and magnetic domains as well as to study magnetic-field induced phase transitions.

To observe phase transitions of the magnetic or crystal structures induced by the magnetic fields, observing the Bragg diffraction or superlattice spots under external magnetic fields is required. Thus, this optical system that is capable of observing Bragg diffraction with the objective lens turned off enables the observation of magnetic field-induced phase transition in a conventional TEM. Note that the phase transition in Fig. 5 can also be observed using the TEM that utilizes a magnetic-field shielding lens as the objective lens and is additionally equipped with an external magnetic-field system. However, our optical system enables us to obtain SmAED patterns with a long camera length in an identical region to that of the Bragg diffraction, and their corresponding Foucault and dark-field images can be obtained only by changing the current in the intermediate lenses, as demonstrated in Fig. 4. Furthermore, this electron optical system can be constructed using any conventional TEMs that contain an objective mini-lens without modifying the mechanical apparatus. In a TEM without the objective mini-lens (HF-3300S, HITACHI Ltd.), we also confirmed that the current optical system can be constructed and that the observation of Bragg diffraction is possible even though the illumination system cannot be controlled to fix the crossover on the SA-aperture plane.

**Concluding Remarks**

We constructed an electron optical system to perform Bragg diffraction with the objective lens turned off. By adjusting the currents in the intermediate lenses, electron diffraction can be observed with a camera length ranging from 0.8 to 4.5 m under the condition that a crossover was located on the selected-area aperture plane. Thus, dark-field images were obtained under controlled external magnetic fields. The diffraction mode was switched to form the SmAED mode by altering the current in the intermediate lenses. We were able to observe the Bragg diffraction pattern and dark-field image in $BaFe_{10.35}Sc_{1.6}Mg_{0.05}O_{19}$ without applying a magnetic field. We also demonstrated using a Ni–Mn–Ga alloy that the crystallographic and magnetic domains can be observed in the same area along with the Bragg diffraction and magnetic deflection patterns using this optical system. We further observed a helical phase transition in the Y-type hexaferrite $Ba_{0.5}Sr_{1.5}Zn_2Fe_{12}O_{22}$ by increasing the magnetic fields. The observation methods disclosed in this study is applicable to many materials exhibiting peculiar magnetic domains and magnetic-field induced phase transitions such as helical magnets, ferromagnetic shape memory alloys, and charge/spin-ordered oxides. Thus, this optical system can play an important role in studying various materials and phase transitions from a microscopic point of view.

**ACKNOWLEDGMENTS**

We are grateful to Prof. Tsuyoshi Kimura of the University of Tokyo for providing the single crystal of the Y-type hexaferrite. We also thank Prof. Yasukazu Murakami of Kyushu University for providing the Ni–Mn–Ga alloy.

**Funding**

This study was partially supported by JSPS KAKENHI (Nos. 16H03833 and 15K13306) and by grants from the Murata Science Foundation.

**Figure Captions**

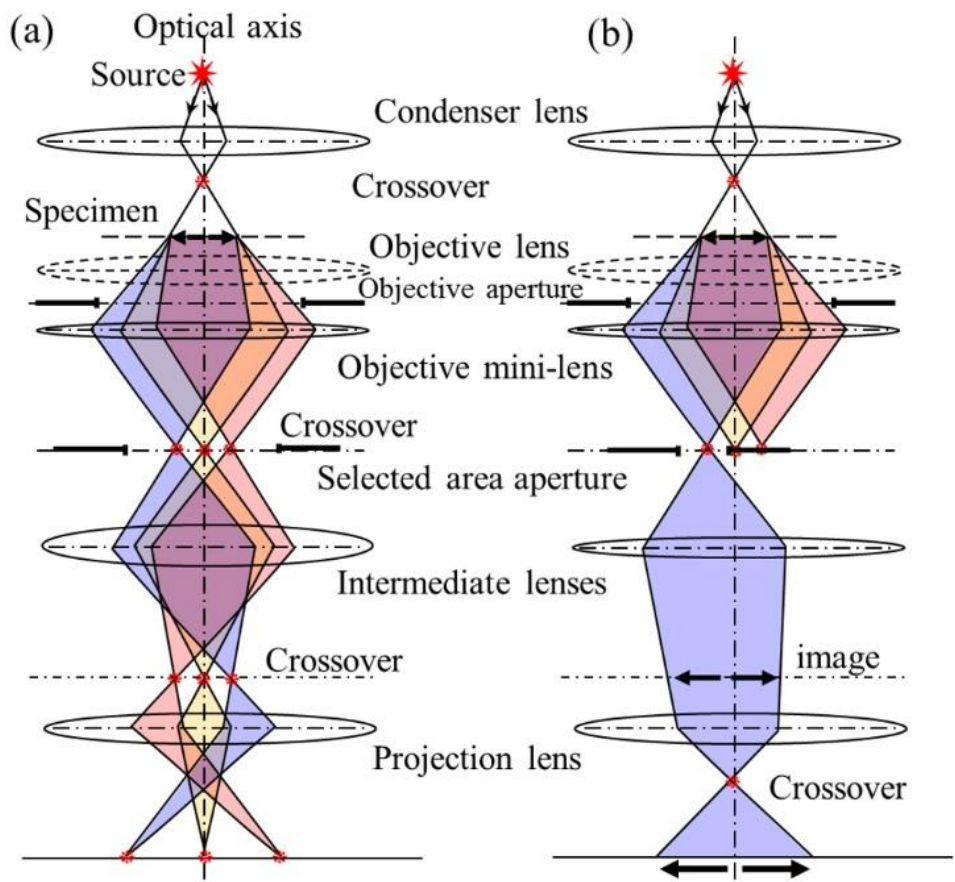

FIG. 1. Schematics of the electron optical systems in (a) electron diffraction and (b) dark-field imaging modes. Intermediate lenses consist of three lenses in this experiment.

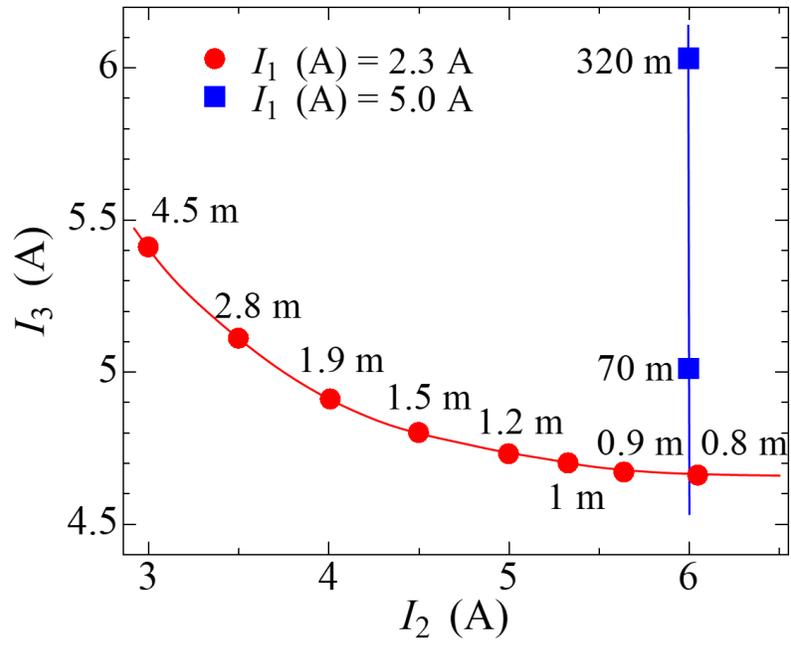

FIG. 2. Camera lengths as functions of the current in the second ($I_2$) and third ($I_3$) intermediate lenses. $I_1$ represents the current in the first intermediate lens. Lines are provided to guide the eye.

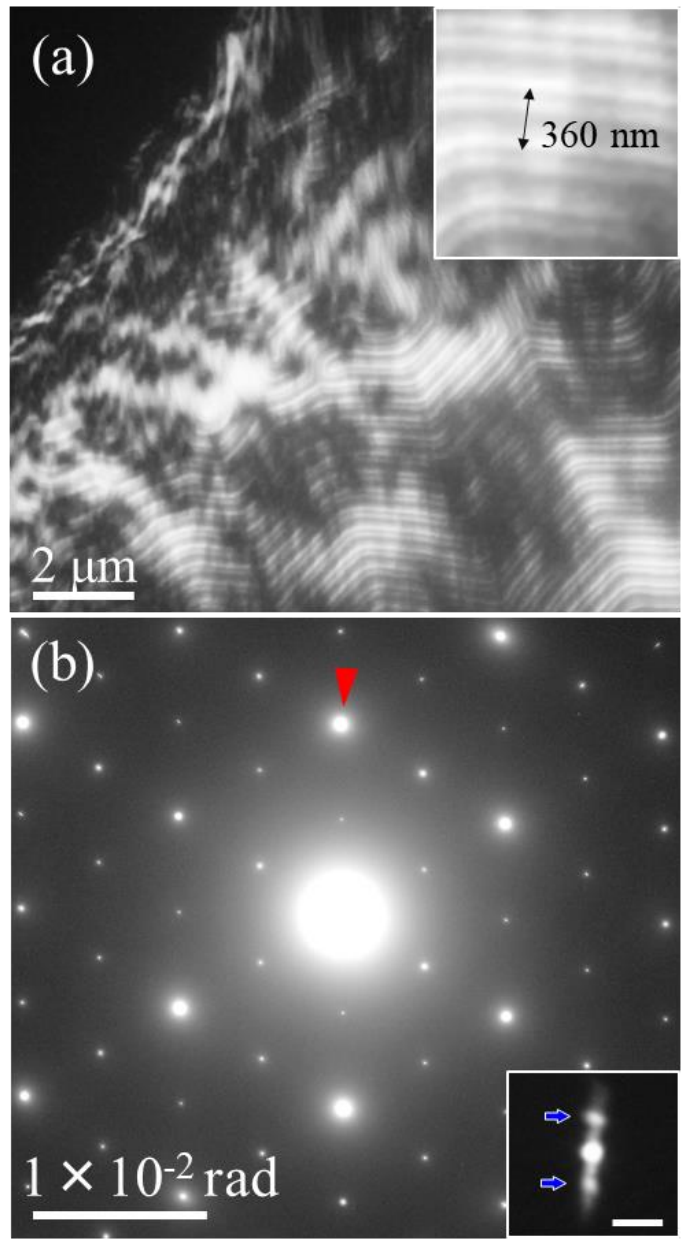

FIG. 3. (a) The dark-field image of BaFe$_{10.35}$Sc$_{1.6}$Mg$_{0.06}$O$_{19}$ using a fundamental reflection 220 indicated by the arrowhead in panel (b). The image was captured at an under-focus condition. The inset depicts a magnified image. (b) Diffraction pattern from the [001]-zone axis at a camera length of 1 m. This diffraction pattern was obtained with the objective lens turned off. The inset depicts an SmAED pattern of the direct beam 000 using a camera length of 320 m. The scale bar is 10 µrad.

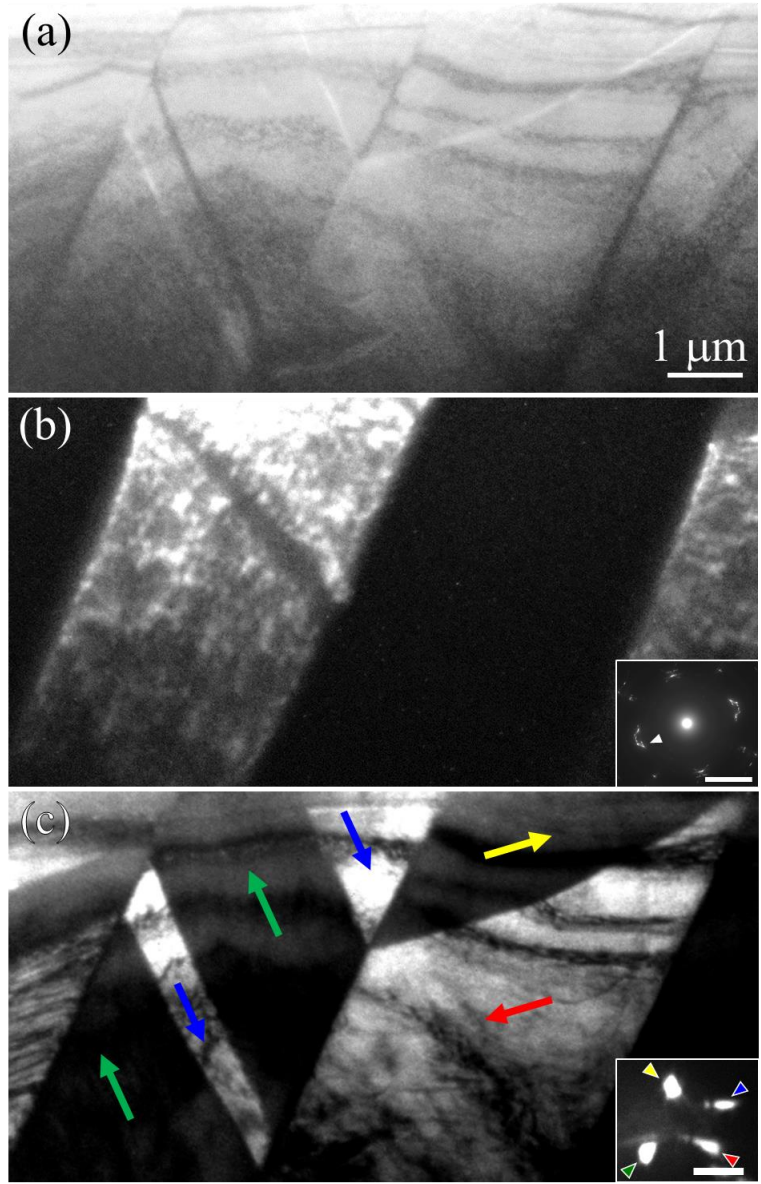

FIG. 4. (a) The underfocused Fresnel image of magnetic domains in the Ni–Mn–Ga alloy at 100 K. (b) The dark-field image of crystallographic twin domains (variants) selecting the Bragg spot indicated by the arrowhead in the inset (Bragg diffraction pattern). This image was obtained at the same area in panel (a). The Bragg diffraction pattern was obtained using a camera length of ~1 m. The scale bar is 10 mrad in the inset. (c) Foucault image for selecting the two magnetic deflection spots that are indicated by red and blue arrowheads in the inset. The arrows illustrate the direction of magnetization. The inset contains an SmAED pattern at a camera length of 100 m using the direct beam 000. The scale bar is 50 μrad in the inset.

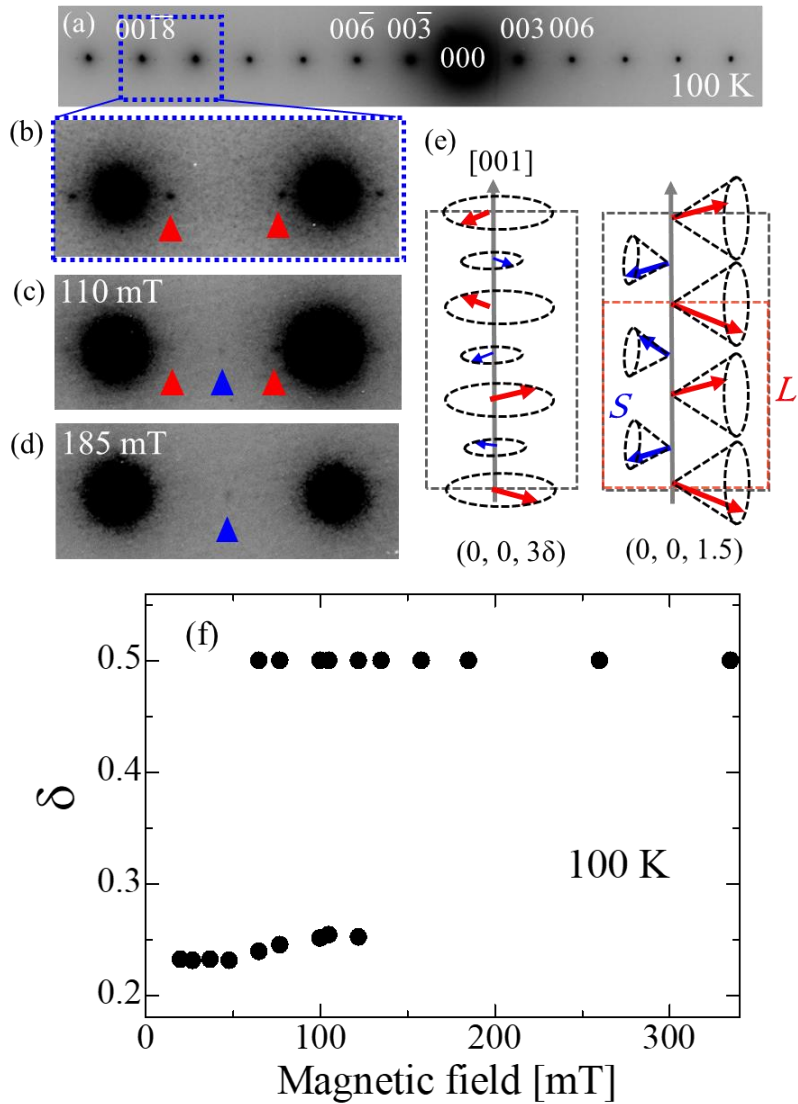

FIG. 5. (a) The electron diffraction pattern of $Ba_{0.5}Sr_{1.5}Zn_2Fe_{12}O_{22}$ at 100 K without the presence of an external magnetic field. The area marked with the rectangular symbol is magnified in panel (b). For clear representation, contrast of the diffraction patterns is reversed. The camera length was ~ 2.0 m. Magnetic-field changes of the superlattice peaks are depicted at (c) 110 and (d) 185 mT. The red and blue arrowheads represent $(0, 0, 3\delta)$ and $(0, 0, 1.5)$ superlattice reflections, respectively. (e) Schematics of the magnetic structures of the $(0, 0, 3\delta)$ proper helical and $(0, 0, 1.5)$ transverse conical structures. The red and blue arrows depict large and small magnetizations in the magnetic structure, respectively. (f) Magnetic-field dependence of the incommensurability $\delta$ at 100 K. The diffraction spots were measured by increasing the magnetic fields.